\documentclass[nonacm,sigplan]{acmart}

\usepackage{amsmath,amsfonts}

\usepackage{graphicx}
\usepackage{textcomp}
\usepackage{braket}
\usepackage{amsthm}

\newtheorem{theorem}{Theorem}[section]
\newtheorem{lemma}[theorem]{Lemma}
\def\BibTeX{{\rm B\kern-.05em{\sc i\kern-.025em b}\kern-.08em
    T\kern-.1667em\lower.7ex\hbox{E}\kern-.125emX}}

\newcommand{\squishlist}{
    \begin{list}{$\bullet$}
        { \setlength{\itemsep}{0pt}      \setlength{\parsep}{0pt}
            \setlength{\topsep}{0.5pt}       \setlength{\partopsep}{0pt}
            \setlength{\listparindent}{-2pt}
            \setlength{\itemindent}{-5pt}
            \setlength{\leftmargin}{0.5em} \setlength{\labelwidth}{0em}
            \setlength{\labelsep}{0.2em} } }
    
\newcommand{\squishend}{
\end{list}  }

\usepackage{tikz}
\newcommand*\circled[1]{\tikz[baseline=(char.base)]{
  \node[shape=circle,draw,fill=black,text=white,font=\bf,inner sep=0.5pt] (char)
  {\small#1};
}}

\newcommand*\circledwhite[1]{\tikz[baseline=(char.base)]{
  \node[shape=circle,draw,fill=white,text=black,font=\bf,inner sep=0.5pt] (char)
  {\small#1};
}}
\usepackage[ruled,vlined]{algorithm2e}

\begin{document}

\title{
Using Reinforcement Learning to Guide Graph State Generation for Photonic Quantum Computers}

\author{Yingheng Li}
\email{yil392@pitt.edu}
\affiliation{%
  \institution{University of Pittsburgh}
  \country{Pittsburgh, USA}
}

\author{Yue Dai}
\email{yud42@pitt.edu}
\affiliation{%
  \institution{University of Pittsburgh}
  \country{Pittsburgh, USA}
}

\author{Aditya Pawar}
\email{adp110@pitt.edu}
\affiliation{%
  \institution{University of Pittsburgh}
  \country{Pittsburgh, USA}
}

\author{Rongchao Dong}
\email{rongchaodong@pitt.edu}
\affiliation{%
  \institution{University of Pittsburgh}
  \country{Pittsburgh, USA}
}

\author{Jun Yang}
\email{juy9@pitt.edu}
\affiliation{%
  \institution{University of Pittsburgh}
  \country{Pittsburgh, USA}
}

\author{Youtao Zhang}
\email{youtao@pitt.edu}
\affiliation{%
  \institution{University of Pittsburgh}
  \country{Pittsburgh, USA}
}

\author{Xulong Tang}
\email{tax6@pitt.edu}
\affiliation{%
  \institution{University of Pittsburgh}
  \country{Pittsburgh, USA}
}

\date{}

\begin{abstract}
Photonic quantum computer (PQC) is an emerging and promising quantum computing paradigm that has gained momentum in recent years. 
  In PQC, which leverages the measurement-based quantum computing (MBQC) model, computations are executed by performing measurements on photons in graph states (i.e., sets of entangled photons) that are generated before measurements. The graph state in PQC is generated deterministically by quantum emitters. The generation process is achieved by applying a sequence of quantum gates to quantum emitters. In this process, i) the time required to complete the process, ii) the number of quantum emitters used, and iii) the number of CZ gates performed between emitters greatly affect the fidelity of the generated graph state. However, prior work for determining the generation sequence only focuses on optimizing the number of quantum emitters. Moreover, identifying the optimal generation sequence has vast search space. To this end, we propose RLGS, a novel compilation framework to identify optimal generation sequences that optimize the three metrics. Experimental results show that RLGS achieves an average reduction in generation time of 31.1\%, 49.6\%, and 57.5\% for small, medium, and large graph states compared to the baseline.
\end{abstract}

\maketitle 
\pagestyle{plain}

\section{Introduction}
Quantum computing has rapidly developed as a promising computing paradigm with significant potential across diverse application domains~\cite{shor,grover,chemistry}. 
Among all types of quantum computers, photonic quantum computers (PQCs) present advantages in fast execution and long coherence time~\cite{photonic, photonicreview}. PQCs leverage photons to represent qubits and apply photon measurements to conduct quantum computations on photonic graph states (i.e., a graph representation of entangled photons)~\cite{stabilizer, hein2006entanglement, mbqc, nielsen2006cluster}.
The photonic graph state is prepared prior to computation. There are two primary methods for generating a photonic graph state: using fusion operations~\cite{fusion} and using quantum emitters~\cite{QDreview, repeater}. While the fusion-based scheme offers advantages in generation speed~\cite{fusion}, it suffers from a low fusion successful rate~\cite{fusion1} where the graph state generation process is not guaranteed to be successful. In contrast, the emitter-based method shows an advantage in \textbf{deterministic} graph state generation, guaranteeing successful graph state generation~\cite{QDreview, repeater, russo2018photonic}. Due to this deterministic advantage, the emitter-based graph state generation scheme has gained increasing interest in recent studies within the physics community~\cite{russo2018photonic, QDreview, zhan2023performance, hilaire2021resource, baseline0}. However, many prior works focus on the fusion-based graph state generation~\cite{oneq, oneperc, mo2024fcm}. The emitter-based approach has received little attention.

To generate a graph state using quantum emitters, a sequence of quantum gates is applied to the emitters, referred to as the ``generation sequence''. Two types of errors are introduced in the generation sequence: decoherence errors and two-qubit gate errors~\cite{hilaire2021resource, russo2018photonic, quantumdot, QDreview}. These errors propagate to the photons within the generated graph state~\cite{russo2018photonic, hilaire2021resource, lindner2009proposal}, leading to reduced accuracy in quantum computations. There are three fidelity metrics associated with these two errors: i) \textbf{Generation time}. The longer the sequence generation time, the higher the decoherence errors~\cite{hilaire2021resource, zhan2023performance, repeater}. Moreover, the long generation time also increases the photon loss rate~\cite{hilaire2021resource, quantumdot, repeater}. ii) \textbf{Number of quantum emitters}. The more number of emitters used in a generation sequence, the higher the decoherence errors~\cite{hilaire2021resource, zhan2023performance, repeater}. iii) \textbf{Number of CZ gates} A greater number of CZ gates used, the worse the two-qubit gate errors~\cite{hilaire2021resource, russo2018photonic, buterakos2017deterministic}.


There are multiple possible generation sequences to produce the same photonic graph state, which consequently affect the aforementioned three fidelity metrics. In this paper, our goal is to identify a graph state generation sequence that collectively optimize all three metrics to enhance the PQC execution fidelity. To the best of our knowledge, the only existing work is Stabilizer Solver~\cite{baseline0}, which primarily optimizes the number of quantum emitters during generation and ignores the other two fidelity metrics. Additionally, the Stabilizer Solver has a vast search space and suffers high complexity in constructing generation sequences (details in Section~\ref{sec: limitation}). 

In this paper, we propose RLGS (\underline{R}einforcement \underline{L}arning-guided \underline{G}raph \underline{S}tate generation), a compilation framework that leverages Reinforcement Learning (RL) and Graph Neural Network (GNN) to efficiently identify the optimal graph state generation sequence. RLGS is built upon graph state operations introduced in~\cite{graph_solver}. 
At its core, RLGS leverages a Q-network to determine the optimal generation sequence. The Q-network is trained offline on various photonic graph states to learn effective strategies. During inference, the trained network constructs the optimal generation sequence for different graph states. RLGS optimizes the three fidelity metrics by designing a reward function to guide the RL agent. 
To further enhance scalability, we employ a ``receptive field'' strategy, where only a subset of photons and emitters is considered at each step, reducing the action space and enabling more efficient decision-making. Compared to the prior approach, RLGS i) takes advantage of the combination of GNN and RL to efficiently solve the graph generation with a vast search space, and ii) collectively optimizes three fidelity metrics and identifies the optimal generation sequence.  The main contributions of this paper are as follows:

\squishlist{}
    \item It identifies three important fidelity metrics involved in photonic graph state generation using quantum emitters. It comprehensively studies these metrics and observes that different graph state generation sequences can have different fidelity results. Thus, it is important to identify an optimal generation sequence that collectively optimizes the three fidelity metrics.   
    
    \item It proposes RLGS (\underline{R}einforcement \underline{L}arning-guided \underline{G}raph \underline{S}tate generation), a novel compilation framework that leverages Reinforcement Learning (RL) and Graph Neural Network (GNN) to efficiently identify the optimal graph state generation sequence. RLGS optimizes the three fidelity metrics by designing a reward function in RL and leverages a ``receptive field'' strategy to reduce the action space and enable efficient decision-making.  
    \item It evaluates RLGS using six representative quantum applications with varying graph state sizes. Compared to the state-of-the-art method, RLGS achieves an average reduction in generation time by 31.1\%, 49.6\%, and 57.5\% for small, medium, and large graph states, respectively. Additionally, the reductions in the number of quantum emitters are 13.9\%, 16.7\%, and 17.5\%, whereas the reductions in the number of CZ gates are 37.7\%, 53.4\%, and 57.8\%.
    
\squishend{}

\section{Background}
\label{sec:intro}

\subsection{Graph State}
\begin{figure}
    \centering
   \includegraphics[width=0.8\linewidth]{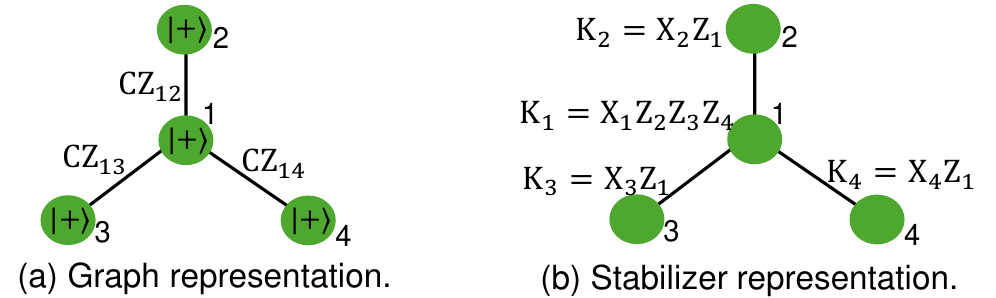}
   \vspace{-10pt}
     \caption{Representations of a graph state.}
    \label{fig:gs}
  \vspace{-15pt}
 \end{figure}

The photonic quantum computers (PQCs) adopt the Measurement-based Quantum Computation (MBQC)~\cite{mbqc, mbqccluster, oneway, nielsen2006cluster} model, where computation is carried out through photon measurements on a group of entangled photons (i.e., physical photon qubits) called graph state~\cite{stabilizer, hein2006entanglement, mbqc, nielsen2006cluster}. 
A graph state $\ket{G}$ has an associated graph structure $G=(V,E)$, where each vertex ($v\in V$) represents a photon (i.e., physical qubit) initialized in the $\ket{+}=(\ket{0}+\ket{1})/\sqrt{2}$ state and each edge ($e\in E$) corresponds to a controlled-Z (CZ) gate between photons. Thus, a graph state can be expressed as:
\begin{equation}\label{eq:graph}
    \ket{G} = \prod_{\{i,j\}\in E}CZ_{ij}\ket{+}
\end{equation}
where $CZ_{ij}$ represent the CZ gate between photon $i$ and photon $j$.
This relationship establishes a direct correspondence between each graph state and a specific graph representation~\cite{mbqc, hein2006entanglement}. An example of the graph representation of a graph state is shown in Fig.~\ref{fig:gs}(a). Additionally, a graph state is also a special type of stabilizer state, where each photon is associated with a Pauli operator~\cite{mbqc, hein2006entanglement, van2004graphical}:

\begin{equation}\label{eq:stabilizer}
    K_i = X_i\prod_{j\in N(i) } Z_j
\end{equation}
where $N(i)$ denotes the set of neighbors of photon $i$, and $X_i$ and $Z_j$ represent the $X$ and $Z$ operator acting on photon $i$ and photon $j$, respectively. Each photon $i$ in a graph state has a unique associated Pauli operator $K_i$ such that $K_i\ket{G} = \ket{G}$, meaning $K_i$ stabilizes $\ket{G}$. The set of these Pauli operators $\{K_i|i\in V\}$ is called the stabilizer of the graph state~\cite{stabilizer, van2004graphical}. 
In Fig.\ref{fig:gs}(b), we annotate the Pauli operator for each photon.
Therefore, a graph state can be represented either as a graph structure or a stabilizer.

\subsection{Graph State Generation}
In general, a graph state is generated before measurements are performed~\cite{mbqc, nielsen2006cluster, quantumdot}. As two-qubit gates are not available between photons~\cite{photonic, photonicreview}, a graph state cannot be directly constructed using CZ gates. Instead, two primary approaches are commonly used to generate graph state: fusion operations and quantum emitters.

\subsubsection{Fusion operations}A photonic graph state can be generated using fusion operations~\cite{fusion}, which merge smaller photonic graph states (also called resource state~\cite{fusion}) into progressively larger ones until the desired photonic graph state is achieved. Although fusion operations offer an advantage in generation speed~\cite{fusion}, they are not resource-efficient for creating photonic graph states, as each fusion operation’s success probability is generally around 75\% to 78\%~\cite{fusion1,78}. Thus, generating the desired graph state involves numerous fusion attempts, leading to a significant number of photons which amplifies other errors (e.g., photon loss)~\cite{oneperc, oneq}. 

\subsubsection{Quantum Emitters}
In contrast, quantum emitters guarantee successful graph state generation ~\cite{previous0, quantumdot, repeater, hybrid}. 
A quantum emitter can emit photons entangled with the emitter itself, functioning as a CNOT gate between the emitter and the emitted photon~\cite{hybrid, previous0}. In addition to the emission gates (i.e., CNOT gates), single-qubit Clifford gates (e.g., H gates) applied to either the emitter or photon, two-qubit gates between emitters, and emitter measurements are utilized in the graph state generation process~\cite{previous0, baseline0, graph_solver}.
We define a sequence of quantum gates applied to quantum emitters to generate a photonic graph state as a ``\textbf{generation sequence}''. 

\subsection{Existing Approach for Generation Sequence}
\label{sec: solver}
The only existing method to identify generation sequences for arbitrary graph states relies on the stabilizer and is referred to as the ``Stabilizer Solver''~\cite{baseline0}.
The Stabilizer Solver is designed to generate photonic graph states using a minimal number of quantum emitters in the generation sequence. Moreover, the algorithm operates in backward order: it begins with the target photonic graph state and disentangled emitters, then determines the sequence of gates to transform this quantum state into a product state (i.e., quantum states with disentangled photons and emitters). After obtaining the backward sequence, the forward generation sequence can be obtained by reversing it~\cite{baseline0, quantum_computation}. The benefit of such backward order is to reduce the number of retries involved in forward order~\cite{baseline0, random_order, graph_solver}. 
Thus, in the rest of the paper, we use backward order as the graph state generation sequence. 

\begin{figure}
    \centering
   \includegraphics[width=1\linewidth]{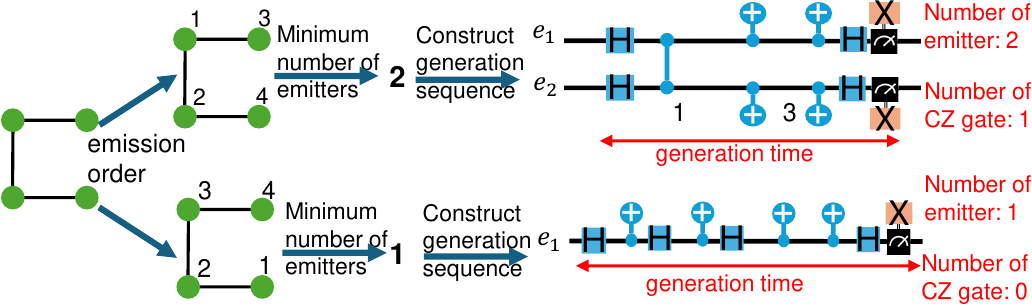}
   \vspace{-15pt}
     \caption{An example of Stabilizer Solver.}
    \label{fig:ss}
    \vspace{-10pt}
 \end{figure}

The Stabilizer Solver relies on a predefined photon emission order. The emission order specifies the sequence in which photons are generated and added to the graph state during the generation process. For a graph state with a specific emission order, the Stabilizer Solver produces a \textbf{single} generation sequence. Given this emission order, the Stabilizer Solver minimizes the number of emitters used to generate the graph state and constructs the corresponding generation sequence. Different emission orders have different generation sequences and may have different values of the number of emitters, the number of CZ gates, and the generation time. For example, in Fig.~\ref{fig:ss}, we show two different emission orders for the same graph state. The numbers next to the photons denote the generation emission order. The first emission order requires a minimum of two emitters to generate the graph state, while the second order requires only one emitter. After determining the minimum number of emitters required, the Stabilizer Solver constructs the corresponding generation sequences~\cite{baseline0}.

\section{Error Model}
\label{sec:error}
Different generation sequences to the same graph state may have different values in graph state fidelity.  
There are two primary types of errors: emitter-caused errors and photon loss.

\subsection{Emitter-Caused Errors} 
\label{sec: emiiter-caused}
Errors occurring in emitters during the generation sequence can propagate to the emitted photons. 
These errors are evenly distributed across all photons in the resulting graph state~\cite{hilaire2021resource, zhan2023performance, lindner2009proposal, buterakos2017deterministic}. There are two types of errors caused by emitters: decoherence errors and CZ gate errors. 

\squishlist{}
    \item \textbf{Decoherence Error}: Decoherence errors occur due to the limited coherence time of the emitters. Each quantum emitter experiences decoherence errors at a rate of $1-e^{\frac{-T_{gen}}{T_2}}$, where $T_2$ represents the coherence time of the emitter and $T_{gen}$ is graph state generation time (i.e.,  the time required to complete a generation sequence)~\cite{quantumdot, hilaire2021resource, zhan2023performance, lindner2009proposal}. If there are $N_e$ emitters within a generation sequence, the fidelity of the emitters affected by decoherence error is given by $F_{de}= e^{\frac{-N_e T_{gen}}{T_2}}$, based on the setting that each emitter remains active throughout the entire generation process~\cite{lindner2009proposal, hilaire2021resource, zhan2023performance, buterakos2017deterministic}. 

    \item  \textbf{CZ gate error}: Emitters face challenges due to the low-fidelity two-qubit gates between emitters. In this paper, we adopt the CZ gate as the native two-qubit gate between emitters~\cite{russo2018photonic, hilaire2021resource, zhan2023performance}. Compared to single-qubit gates on emitters, CZ gates between emitters introduce a significantly higher error rate~\cite{QDreview, hilaire2021resource, russo2018photonic, random_order, buterakos2017deterministic}. The fidelity for emitters affected by CZ gate errors is $F_{CZ} =\sigma_{CZ}^{N_{CZ}}$, where $\sigma_{CZ}$ is the fidelity of a single CZ gate and $N_{CZ}$ is the total number of CZ gates used in a generation sequence~\cite{russo2018photonic, hilaire2021resource}.  
    
\squishend{}
As such, both emitter-caused errors are affected by i) \textbf{the graph state generation time} ($T_{gen}$), ii) \textbf{the number of emitters} ($N_e$), and iii) \textbf{the number of CZ gates} ($N_{CZ}$). For example, in the first emission order in Fig.~\ref{fig:ss}, the generation sequence requires $N_{e}=2$ and $N_{CZ}=1$, with $T_{gen}$ including the time required to execute two H gates, one CZ gate, and two CNOT (i.e., emission) gates. For the second generation sequence, it requires $N_{e}=1$ and $N_{CZ}=0$, with $T_{gen}$ including the time required to execute four H gates and four CNOT gates. 

\subsection{Photon Loss}
\label{sec:loss}
Photon loss is another type of error in PQCs~\cite{hilaire2021resource, zhan2023performance, quantumdot, repeater}. 
Photon loss can occur in both the graph state generation stage and the actual measurement computation stage. In the graph state generation stage, the generated photons are stored in delay lines (e.g., optical fiber)~\cite{quantumdot,hilaire2021resource, zhan2023performance, oneperc}. The longer the photon is stored in delay lines, the higher the probability of loss~\cite{fiber, quantumdot}. Since all the generated photons remain in the delay lines before the actual measurement computation stage, the photon loss rate in the graph state generation stage is affected by the generation time $T_{gen}$ (i.e., time to keep the first photon in delay line till all photons are generated and the measurement computation stage can start). 
The loss in an optical fiber is typically expressed as $L\frac{dB}{km}$, meaning the probability of a photon remaining in the optical fiber (i.e., no loss) after traveling a distance of $d$(km) is $P_{remain}=10^{\frac{-L\cdot d}{10}}$~\cite{fiber}. Given the light speed in the optical fiber is 200,000$km/s$~\cite{wiki:Optical_fiber}, the probability of a photon remaining in the fiber for a generation time $T_{gen}$(in nanoseconds) can be expressed as: $P_{remain}=10^{\frac{-L\cdot T_{gen}}{50,000}}$. 



It is important to emphasize that the photon loss rate during the actual measurement computation stage is much higher than that during the graph state generation stage 
(e.g., 10 orders of magnitude higher)~\cite{quantumdot}. Mitigating the photon loss rate in measurement computation is an active research field~\cite{photonicreview, zhan2023performance, hilaire2021resource, hybrid} and beyond the scope of this paper.

\section{Motivation and Opportunity}


\begin{figure*}
    \centering
   \includegraphics[width=0.9\linewidth]{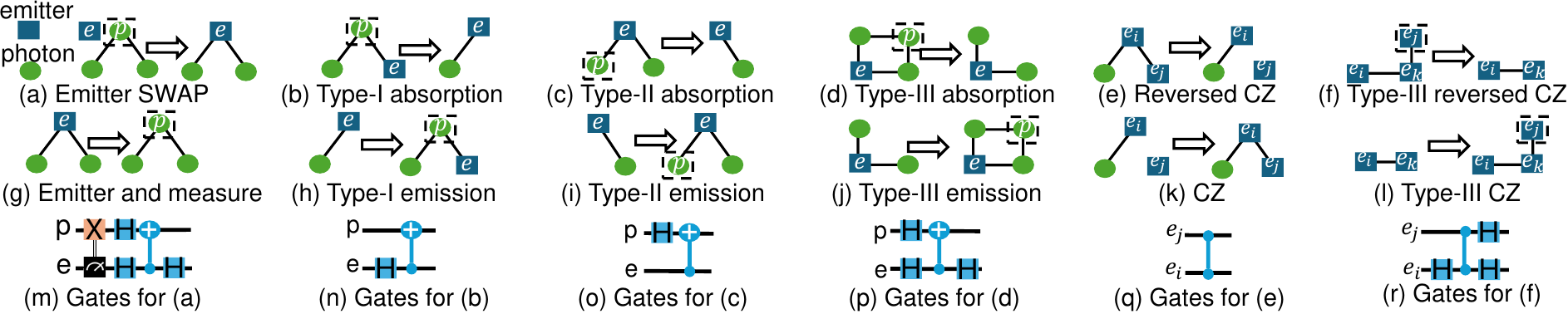}
   \vspace{-7pt}
    \caption{Examples of graph operations, their forward versions, and their quantum gate.}
    \label{fig:six}
    \vspace{-10pt}
 \end{figure*}

\subsection{Limitation of the Stabilizer Solver}
\label{sec: limitation}
Recall that the Stabilizer Solver is the only existing method to generate graph state using photon emitters. There are two limitations of the Stabilizer Solver:

\squishlist{}
    \item \textbf{Ignorance of generation time and number of CZ gates}: The Stabilizer Solver minimizes the number of emitters for a given emission order, overlooking other two metrics: the graph state generation time $T_{gen}$ and the number of CZ gates $N_{CZ}$. 
    However, as we discussed in Section~\ref{sec:error}, all three metrics play an important role in the related errors involved in graph state generation. Therefore, it is important to collectively consider all three metrics when finding a good graph state generation sequence.      
    \item \textbf{Large search space and high complexity}: Recall that the Stabilizer Solver relies on a given emission order to construct the corresponding graph state generation sequence. Assuming there are $V$ photons in the target graph state, there are a total of  $V!$ possible emission orders. For the Stabilizer Solver to find the optimal generation sequence, it needs to exhaustively iterate through all the $V!$ emission orders, leading to a large search space. Additionally, constructing the generation sequence for each emission order has a complexity of $O(V^4)$~\cite{baseline0}. Therefore, the overall complexity of constructing an optimal generation sequence for a graph state using the Stabilizer Solver is $O(V! \cdot V^4)$, which becomes very high when the graph state size increases. 

\squishend{}




\begin{figure}
    \centering
   \includegraphics[width=1\linewidth]{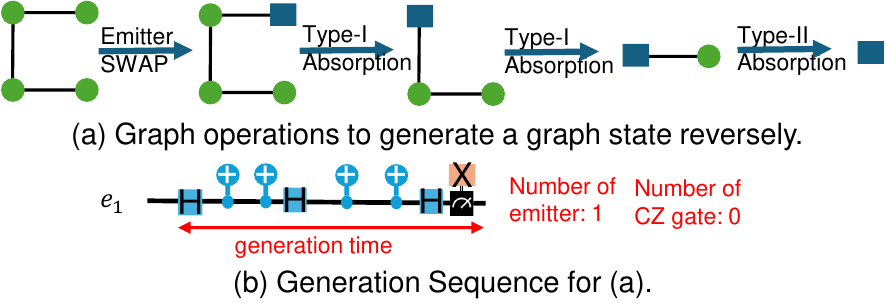}
   \vspace{-20pt}
     \caption{An example of using graph operations.}
    \label{fig:gs_example}
    \vspace{-16pt}
 \end{figure}
 
\subsection{Graph Operations} 
\label{sec:graph solver}
Recent work proposed six graph operations to efficiently generate graph state using emitters~\cite{graph_solver}. The six graph operations avoid the overheads of constructing an optimal generation sequence in the Stabilizer Solver. 
The six operations are also present in a backward order. 
We illustrate the six graph operations in Fig.~\ref{fig:six}(a)-(f), with the removed photons and emitter highlighted in a dotted square. The corresponding forward versions of these graph operations are shown in Fig.\ref{fig:six}(g)-(l), and the associated quantum gates for each graph operation are presented in Fig.~\ref{fig:six}(m)-(r). Specifically: 


\squishlist{}
    \item \textbf{Emitter Swap} (Fig.~\ref{fig:six}(a)): A newly-initialized (i.e., disentangled) emitter $e$ replaces the $p$ photon in a graph state $\ket{G}$. Any photon in a graph state can be replaced by a newly-initialized emitter. 
    

    \item  \textbf{Type-I absorption} (Fig.~\ref{fig:six}(b)): If a graph state $\ket{G}$ includes an emitter $e$ and a photon $p$ with $N(e)=\{p\}$ (where $N$ denotes the set of neighbors), the photon $p$ can be absorbed by the emitter $e$. In other words, Type-I absorption occurs when the neighbor of an emitter consists of only a single photon. After performing a Type-I absorption, $e$ replaces $p$ in the graph structure.    
    
    \item \textbf{Type-II absorption} (Fig.~\ref{fig:six}(c)): If a graph state $\ket{G}$ contains an emitter $e$ and a photon $p$ with $N(p)=\{e\}$ (i.e., the photon has only one neighbor and the neighbor is an emitter), the emitter $e$ can absorb photon $p$, removing the edge between them as well. 

    \item \textbf{Type-III absorption} (Fig.~\ref{fig:six}(d)): Type-III absorption can occur between an emitter $e$ and a photon $p$ if $N(p)=N(e)$, meaning the emitter $e$ and the photon $p$ have the same set of neighbor nodes. Type-III absorption removes $p$ along with its associated edges from the graph. 

    \item \textbf{Reversed CZ} (Fig.~\ref{fig:six}(e)): The reversed CZ removes an edge between two emitters $e_i$ and $e_j$. Note that, according to \cite{baseline0, graph_solver, random_order}, CZ gates can be applied to any two emitters regardless of their distances~\cite{long1, long2}.
    \item \textbf{Type-III reversed CZ} (Fig.~\ref{fig:six}(f)):
    Similar to the Type-III absorption, a Type-III reversed CZ can be applied between two emitters $e_i$ and $e_j$ in a graph state if $N(e_i)=N(e_j)$~\cite{baseline0, graph_solver}. This operation removes $e_j$ along with its associated edges from the graph state.

    
\squishend{}

We provide an example in Fig.~\ref{fig:gs_example}(a), demonstrating how graph operations are used to construct a generation sequence in backward order for the same graph state from Fig.~\ref{fig:ss}. Initially, assuming photon 4 is swapped with an emitter using the Emitter SWAP operation. Next, a Type-I absorption is applied to photon 3, followed by another Type-I absorption on photon 2. Finally, a Type-II absorption is applied to photon 1. Once all edges and photons are removed, the process concludes. The constructed generation sequence in forward order is shown in Fig.~\ref{fig:gs_example}(b). 



Using graph operations to construct a generation sequence has two advantages over the Stabilizer Solver. First, graph operations give flexibility during generation by providing six operations. This allows the generation sequence to be optimized for different fidelity metrics, while the Stabilizer Solver only minimizes the number of emitters~\cite{graph_solver}. 
Second, it avoids constructing a generation sequence in Stabilizer Solve which has  $O(V^4)$ complexity. Since a graph operation either removes a photon or an edge, constructing a generation sequence requires at most $V+E$ operations, where $V$ is the number of photons and $E$ is the number of edges in a graph state, with only six choices per operation. Therefore, the complexity of constructing a generation sequence is reduced to $O(V+E)$.
Therefore, in this paper, we leverage graph operations to generate graph states. 


\subsection{Opportunity}
\label{sec:opportunity}

Despite the advantages of using graph operations mentioned above, it is still challenging to use graph operations to generate a graph state which i) collectively optimizes the three fidelity metrics and ii) efficiently finds the optimized generation sequence in a large search space. On the one hand, each graph operation may affect the three fidelity metrics, and one needs to carefully choose the operation during the generation process. On the other hand, to find the optimal generation sequence, one needs to exhaustively search all the generation sequences (i.e., $V!$) as in Stabilizer Solver (Section~\ref{sec: solver}). To efficiently determine the generation sequence by leveraging the graph representation provided by the graph operations, we propose to take advantage of two powerful tools: Graph Neural Networks (GNN) and Reinforcement Learning (RL). We use GNN to extract the structural relationships of a graph state and use RL to learn and guide the generation sequence construction. 

\subsubsection{Graph Neural Network}

GNNs extend neural network architectures to process graph-structured data \cite{kipf2016semi,hamilton2017inductive,velivckovic2017graph,xu2018powerful}, enabling encoding the local neighborhood information around each node into its feature representation, which can be leveraged by downstream tasks to solve various graph-related problems~\cite{bai2020learning,bai2021glsearch,dai2023cegma,wang2021combinatorial}.
A typical GNN consists of multiple layers, where each layer allows nodes to gather and process information from their 1-hop neighbors, producing latent representations in two steps as follows.
\begin{equation}\label{eq:gnn_layer}
    X^{l+1} = \sigma(Comb(Aggr(A,X^{l},W^{l}_{e}),W^{l}_{n}))
\end{equation}
where $\sigma(\cdot)$ is the activation function, $A$ is the adjacency matrix of the graph, $X^{l}$ is the node feature at layer $l$, and $W^{l}_{e}$ and $W^{l}_{n}$ are the weights of layer $l$. 
The aggregation function $Aggr(\cdot)$ collects messages from a node's neighbors, typically implemented as max, average, or mean pooling \cite{hamilton2017inductive} or as attention-based weighted sums \cite{velivckovic2017graph}. The combination function $Comb(\cdot)$ then update nodes' embedding on top of the aggregated messages.
With multiple layers in GNNs, nodes iteratively aggregate and combine information from their 1-hop neighbors. Consequently, after $K$ layers, the node embeddings reflect information from a $K$-hop subgraph \cite{kipf2016semi,hamilton2017inductive,velivckovic2017graph,xu2018powerful}.
Lastly, one can use a readout function, such as average pooling, to retrieve the embedding of a graph from its node embeddings.

\subsubsection{Reinforcement Learning}
Building on the success of GNNs, recent studies have integrated Reinforcement Learning (RL) to tackle NP-hard problems on graphs~\cite{bai2021glsearch,wang2021combinatorial,munikoti2023challenges,ma2019combinatorial}. These methods typically utilize graph embeddings generated by GNNs to inform decision-making processes within RL agents. These methods typically employ Deep Q-Networks (DQN)~\cite{DQN0} as the RL agent.
DQN is a simple yet effective approach that has been successfully applied to various real-world decision-making problems~\cite{DQN1, DQN0, bai2021glsearch, munikoti2023challenges}. 
Generally, during each decision-making step, 
the RL agent estimates the expected reward for each choice and selects the one that maximizes this reward. This reward typically represents a value that describes the best possible outcome for that choice, such as the shortest path length achievable by selecting a specific neighbor in the shortest path problem.
By leveraging these estimates, decisions can be made based on their potential outcomes without the need to explicitly explore them, hence the search costs are significantly reduced.
To achieve this, one needs to train a Q-network $Q_{\theta}(s,a)$ so that it can estimate the optimal Q-function value $Q^*(s,a)$, which represents the maximum cumulative reward starting from state $s$, taking action $a$, and following the best possible future actions.

\begin{equation}\label{eq:q_function}
    Q^{*}(s,a) = \mathbb{E}[r+\gamma \max_{a'}Q^{*}(s',a')]
\end{equation}
$Q^*(s,a)$ is calculated by the sum of immediate reward $r$ and the expected future reward, discounted by $\gamma$. The return is obtained by following the optimal actions until the end (i.e., the maximum reward from the next state $s'$ and following the best action $a'$). 
By iteratively training $Q_{\theta}(s, a)$, the Q-network learns to predict $Q^*(s, a)$, making it possible to make decisions that maximize long-term rewards without exhaustively exploring all possible outcomes.


The combination of GNN and RL(DQN) provides several key advantages in solving our problem:

\squishlist{}
    \item \textbf{Efficiently exploring vast search space}: As discussed above, the search space for constructing a generation sequence is vast, making it challenging to identify the optimal sequence within this space. The combination of GNN and RL has been shown to efficiently handle huge search space in graph-related problems.
    \item \textbf{Collectively metrics optimization}: 
    RL enables simultaneous optimization of all three metrics by defining an appropriate reward function, guiding its action selection with the specified optimization objectives (i.e., a combination of the three metrics)~\cite{DQN0, DQN1}.
    \item \textbf{Long-term decision-making}: RL is well-suited for problems where actions have long-term effects on future states
    ~\cite{DQN0, DQN1, bai2021glsearch}.
    This capability allows the model to consider both immediate outcomes and expected future results, whereas a heuristic is limited to focusing only on immediate outcomes. In our case, early graph operation choices significantly impact subsequent ones, making RL ideal for identifying optimal operation sequences in this context.
    \item \textbf{Generalization}: We observe that many graph states share a similar pattern, allowing a Q-network to learn these patterns and generalize them to solve unseen patterns in different graph states. In our approach, we train a single Q-network using a representative set of graph states and apply it to solve other graph states. This strategy reduces computational overhead by eliminating the need for retraining on each instance and enhances the generalization.  

\squishend

\section{RLGS Design}
In this paper, we design RLGS (\underline{R}einforcement \underline{L}earning-guided \underline{G}raph \underline{S}tate generation), a compilation framework that leverages RL to identify the optimal generation sequence for constructing graph states using emitters.
Our proposed RLGS is achieved through two stages: The \textbf{offline training} and \textbf{inference}, as shown in Fig.~\ref{fig:RLGS}.

During the offline training stage, two phases are performed repeatedly: \textit{experience} and \textit{model training}. 
In the experience phase, the RL agent explores the generation sequences of graph states in the training set and records the outcomes of each exploration. 
In the model training phase, the recorded data is used to train the Q-network.
The details of the are included in Section~\ref{subsec: training}.

During the inference stage, RLGS takes a photonic graph state as input and uses the trained Q-network to determine the best graph operation sequence, as detailed in Section~\ref{subsec: inference}. 
In particular, given a graph state, it leverages GNNs to embed its graph information, then uses the trained Q-network to predict the outcome of each operation and, lastly, selects the one that gives the optimal outcome.

\subsection{RL Formalization}
\label{sec:formulation}
To begin, we first introduce how RLGS formalizes the problem of identifying the optimal graph state generation sequence as an RL problem. 
Recalling our discussion in Section~\ref{sec:opportunity}, there are three key components that are crucial for RL: State, Action, and Reward.
Overall, in RLGS, we define the problem as:
At each graph state (as \textbf{\textit{state}}), the RLGS uses a Q-network to predict the combination of cumulative fidelity metrics (the combinations as \textbf{\textit{reward}} and their cumulative value as \textbf{\textit{Q-value}}) of each graph operation (as \textbf{\textit{action}}) and then selects the one that yields the optimal metrics.
Details of the components are as follows:

\subsubsection{State} 
In RLGS, given a specific iteration, we use a state to describe all remaining photons, emitters, and their entanglements (i.e., those not yet absorbed or removed).
A state is defined as a graph $G=(V,E)$, where $V$ is the set of vertices, with each vertex representing either a photon or an emitter. $E$ is the set of edges, where each edge corresponds to an entanglement between photons or emitters. 
Given the state at iteration $t$, RLGS utilizes a GNN to encode its information into a state vector $s_t$ (e.g., 128-dimension vector) for the following decision-making process.
The initial input state $s_0$ for RLGS is a photonic graph state where every vertex is a photon and all photons are entangled. 


\subsubsection{Action} An action $a_t$ in RLGS refers to one of the six graph operations introduced in Section~\ref{sec: solver}, applied to one of the vertexes or between two vertexes in a state. 
After applying an action $a_t$ to a state $s_t$, a new state $s_{t+1}$ is generated. Note that each combination of a state and action produces a unique new state because emitter operations are deterministic~\cite{graph_solver, baseline0}. 
We only consider the actions that satisfy the operation constraints introduced in Section~\ref{sec:graph solver}.
\subsubsection{Reward and Q-value} The reward in RLGS represents a combination of the three metrics: generation time $T_{gen}$, emitter number $N_e$, and CZ gates number $N_{CZ}$. To reduce $T_{gen}$, we define the reward function as the negative value of the additional generation time required to execute a given operation. The reward is set to a negative value since our objective is to minimize the total generation time. Additionally, we can reduce both $N_e$ and $N_{CZ}$ by applying a penalty for each newly introduced emitter through Emitter Swapping because $N_{CZ}$ is proportional to $N_e$ ($N_{CZ} \propto N_e$), as detailed in \textit{Lemma}~\ref{theo: 1}.

\begin{lemma}\label{theo: 1}
For a graph state with $V$ photons and $E$ edges, each photon can either be swapped or absorbed by an emitter. Therefore, the number of photon absorption operations is $N_{absor}=V-N_e$. Edges in the graph state can be removed through photon absorption or CZ gates. Each operation can remove one edge
or multiple edges.
Assuming the average edges removed by a photon absorption is $\beta$ and the average edges removed by a CZ gate is $\gamma$, the relationship between the number of edges and operations can be expressed as $E = \beta N_{absor} + \gamma N_{CZ}$.
Rearranging, we have: $N_{CZ} = \frac{E}{\gamma} - \frac{\beta}{\gamma}V + \frac{\beta}{\gamma}N_{e}$. Since $E$ and $V$ are constant for a given graph state, $N_{CZ} \propto N_e$.
\end{lemma}

To this end, we set the reward function in RLGS as:

    \begin{equation}\label{eq:reward}
    r(s_t, a_t) = \begin{cases}
    -add(T_{gen})- \alpha \cdot T_{CZ},& \text{if $a_t=SWAP$}\\
    -add(T_{gen}),  & \text{otherwise}
\end{cases}
\end{equation}
where $add(T_{gen})$ represents the additional generation time after applying an action, $\alpha$ is a user-defined parameter, and $T_{CZ}$ is a constant time required to operate a CZ gate. If the action involves an Emitter Swap, the reward is calculated as the additional time to execute the action plus a fraction of the CZ gate time. 
As discussed in~\cite{QDreview, russo2018photonic, zhan2023performance}, different emitters exhibit varying characteristics.
The user can adjust $\alpha$ to align with the specific requirements of the selected emitter, placing greater emphasis on $T_{gen}$ or prioritizing the other two metrics.
In RLGS, the Q-value represents the predicted cumulative future reward from the current state to the end state by taking the optimal actions.

\subsection{Offline Training of RLGS}
\label{subsec: training}
Before applying RLGS to construct generation sequences for diverse graph states, its Q-network is trained offline using a specific set of training data (e.g., small graph states). The objective of the training is to enable the Q-network to accurately predict the reward of performing a graph operation for any given graph state.
Once the Q-network is trained, it can be applied to different photonic graph states (i.e., inference). 

Before the training, RLGS sets up a replay buffer $D$ of size $M$ for training, initializes a Q-network with random weights $\theta$, and creates a target Q-network with weights identical to the main Q-network. The RLGS training process is repeated for $N$ episodes, where each episode involves identifying a generation sequence for a given photonic graph state. 
Each training episode consists of two repeatedly conducted phases: the \textit{Experience phase} and the \textit{Model Training phase}. 
During the \textit{Experience phase}, RLGS uses its current Q-network to construct generation sequences and record the information during this process. 
During the \textit{Model Training phase}, the recorded transitions are used as ground truth to guide the training of the Q-network. 
The \textit{Model Training phase} is triggered after each decision (i.e., once a graph operation is selected) in the \textit{Experience phase}.

\begin{figure}
    \centering
   \includegraphics[width=1\linewidth]{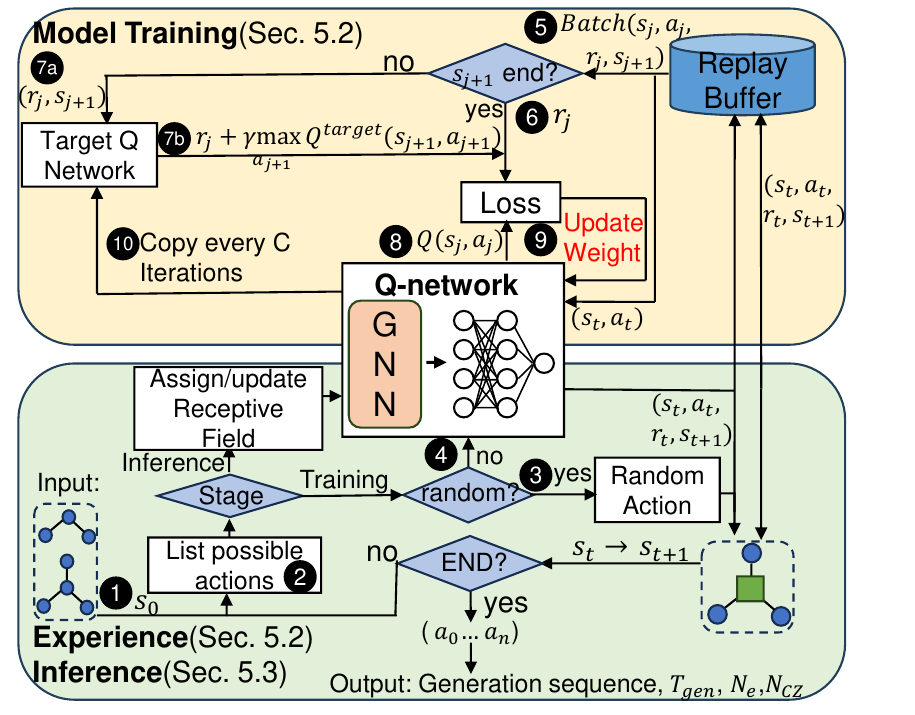}
   \vspace{-15pt}
     \caption{The overview of RLGS.}
    \label{fig:RLGS}
 \vspace{-20pt}
 \end{figure}

\textbf{Experience Phase.} We illustrate the procedure of the experience phase in Fig.\ref{fig:RLGS}. RLGS begins by setting the initial state $s_0$ as the input graph state $G$ (step \circled{1}). 
Next, given the state (generally denoted as $s_t$), RLGS finds all applicable graph operations as possible actions and generates a list of all action-next state pairs, denoted as $\{(a_t, s_{t+1}), \dots\}$ (step \circled{2}).
Then, it decides whether it is conducting inference or experience phase.
If it is the experience phase, the RLGS will use a $\epsilon$-greedy policy to decide whether to (a) conduct a random action from the list (step \circled{3}) or (b) take the action that maximizes the Q-value.
As for the latter choice, the Q-network $Q$ is used to select the action-next state pair that is expected to maximize future rewards (i.e., the action that yields the highest Q-value)(step \circled{4}). 
After obtaining the immediate reward $r_t$, RLGS updates the current state to the new state $s_{t+1}$ and stores related information in replay buffer $D$ for training. Specifically, given an action applied at iteration $t$, the information, referred to as \textit{\textbf{transition}} in the following text, includes the action performed (i.e., $a_t$), the graph states before and after the action (i.e., $s_t$ and $s_{t+1}$), and the corresponding reward $r_t$ (a combination of fidelity metrics).

Regarding the $\epsilon$-greedy policy mentioned above, we use a hyper-parameter $\epsilon$ to control the probability of randomly selecting actions~\cite{DQN0,DQN1}. That is, with $\epsilon$ probability, the agent will select a random action (i.e., explore) instead of applying the predicted best choice (i.e., exploit). 
This random exploration ensures that the RLGS can occasionally check other options instead of trusting its current knowledge, avoiding getting stuck in local optima.
In the early episodes, $\epsilon$ is set high, giving RLGS a greater likelihood of selecting random actions to encourage exploration. Exploration helps the agent avoid local optima by discovering new actions and states that a trained Q-network might overlook if it focused only on familiar actions. After each episode, $\epsilon$ is updated to $\epsilon \cdot \delta$, where $\delta<1$, gradually reducing the exploration rate. As training progresses and $\epsilon$ decreases, RLGS shifts toward exploitation. Exploitation allows the Q-network to converge by consistently selecting the best actions, reinforcing optimal decision-making.

\textbf{Model Training Phase.} In the training phase, a batch of $B$ transitions is sampled from the replay buffer (step \circled{5}).
Then, RLGS computes the ground truth (i.e., Q-values) for the Q-network. This is done in two cases:
(a) If $s_{j+1}$ is an end state, the target Q-value $y_j$ is simply the immediate reward $r_j$  (step \circled{6}). 
(b) If $s_{j+1}$ is not the end state, the Q-value is computed in two steps: RLGS first uses a target Q-network, which is a stale copy of the Q-network, to predict the Q-values of taking any possible actions in the next state (i.e., Q-value of taking $a_{j+1}$ on top of $s_{j+1}$ (step \circled{7a}). Then RLGS selects the maximum Q-value among all potential subsequent actions (i.e., $max(Q^{\text{target}}(s_{j+1},a_{j+1})$), discount it by $gamma$ and add the reward of the current action $a_j$ (i.e., $r_j$) to simulate Q-value of performing $a_j$ at state $s_j$ (step \circled{7b}).
With the ground truth determined, RLGS proceeds to train the main Q-network: 
RLGS uses the Q-network to predict the Q-value $Q(s_j,a_j)$ based on the current graph state $s_j$ and action $a_j$ (step \circled{8}). Then, RLGS calculates the loss by comparing the predicted Q-value with the ground truth and uses the error to update the parameter $\theta$ in the Q network (step \circled{9}).
Every $C$ iteration, the weights of the target Q-network $Q^{\text{target}}$ are synchronized with those of the main Q-network $Q$ (step \circled{9}). 

Instead of using the main Q-network, RLGS uses a target Q-network $Q^{\text{target}}$ to stabilize the training process~\cite{DQN0, DQN1}. Without it, the Q-values can become unstable because the main Q-network would be changed continuously due to its updated parameters. By periodically updating $Q^{\text{target}}$ (step \circled{10}), the learning process is stabilized, as it provides a relatively fixed reference point for Q-network updates~\cite{DQN0, DQN1}.

 \begin{figure*}
    \centerline{\includegraphics[width=0.9\linewidth]{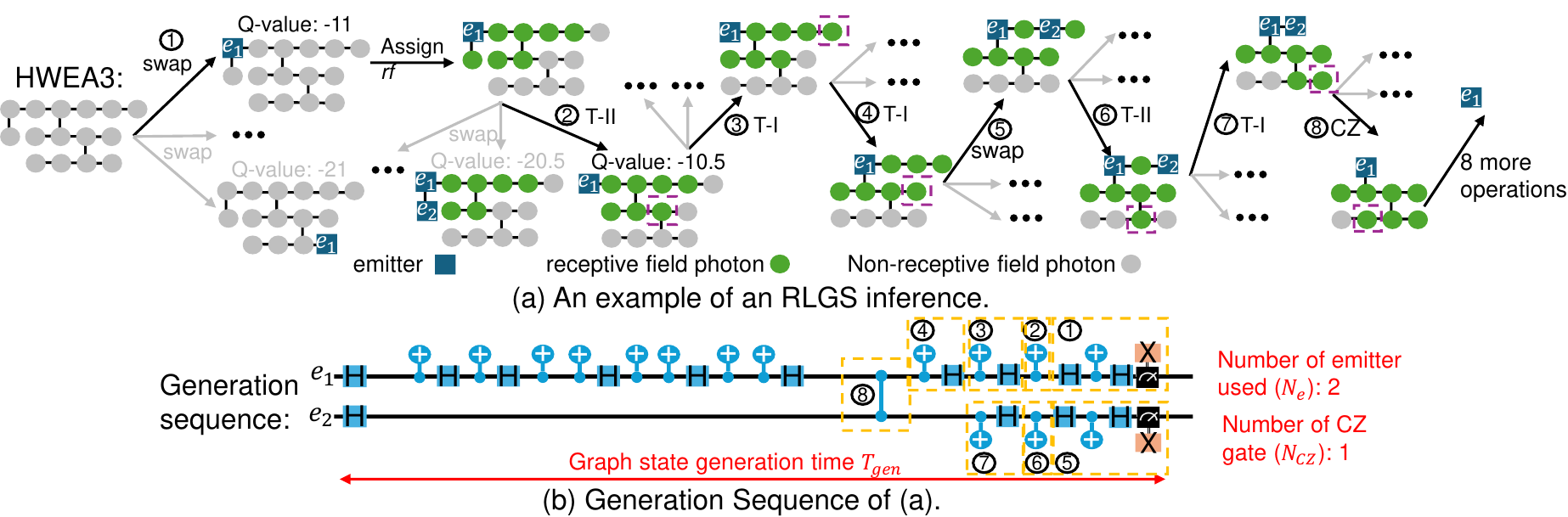}}
    \vspace{-6pt}
    \caption{An example of an RLGS inference stage.}
    \label{fig:example}
\vspace{-10pt}
\end{figure*}

\subsection{Inference of RLGS}
\label{subsec: inference}
Once a trained Q-network is obtained, RLGS uses it to determine an action sequence that identifies a generation sequence that yields optimal fidelity metrics for a given photonic graph state.
As shown in Fig.~\ref{fig:RLGS}, the inference process is similar to the \textit{Experience Phase} in the model training. Specifically, RLGS determines graph operations by selecting the actions with the highest predicted Q-value until all photons and edges are removed.

Before inferencing, RLGS initializes the state $s_0$ as the input photonic graph states $G$. Moreover, it also initialized an empty action sequence $A$, an empty distance table, and an empty receptive field $rf$. 
At the first step, RLGS begins by using the Q-network to decide an optimal Emitter Swap operation, as the initial state only consists of photons. 
Once the best action-next state pair $(a_t^*, s_{t+1}^*)$ is identified and the initial photon is recognized, RLGS adds the action $a_0^*$ to the action sequence $A$, and the current state is updated to $s_{1}^*$. 
Moreover, based on the initial photon, RLGS assigns vertices to the receptive field $rf$ in the following two steps:
First, RLGS computes the shortest path distances (i.e., Manhattan distances) between the initially assigned emitter and the remaining photons, then sorts the photons based on their distance to the emitter and stores the sorted indices in the distance table.
Next, RLGS assigns the initially assigned emitter and the $W-1$ closest photons (i.e., represented by the first $W-1$ entries in the distance table) to the receptive field. 
In the following inference, RLGS only considers the graph operations on vertices inside the receptive field. 
By doing so, the Q-network only considers actions involving up to $W$ vertices in each iteration. This reduces the action space and consequently reduces the complexity of RLGS inference, providing better scalability.


In each subsequent iteration, RLGS continues to generate subsequent graph operations one by one. For each iteration, it determines the best graph operation in three steps:
(a) First, RLGS recognizes actions that are applicable to vertices within the receptive field $rf$. Specifically, for each photon or emitter, RLGS finds the applicable graph operations and treats each of them as a potential action.
(b) Second, the Q-network then selects the optimal action–next state pair $(a_t^*, s_{t+1}^*)$ that maximizes the Q-value. The chosen action $a_t^*$ is added to the action sequence $A$, and the current state $s_t$ is updated to $s_{t+1}^*$. 
(c) Once a vertex is removed from the receptive field (e.g., absorbed by an operation), RLGS maintains its size by adding the closest photon outside the receptive field to the initially assigned emitter based on the distance table.
The inference process ends when the current state $s_t$ has no remaining edges and photons. The action sequence $A$ is then returned as the output, representing the optimal sequence of graph operations.

\subsection{Running Example of RLGS Inference}\label{subsec: example}
We show an example of the RLGS inference stage in Fig.~\ref{fig:example}(a), where RLGS is used to find the generation sequence for a 3-logical-qubit hardware efficient ansatz ({\tt hwea-3})~\cite{hwea} photonic graph state.

As shown in step \circledwhite{1}, in the beginning, the Q-network evaluates all actions to determine the first photon to be swapped. With 15 photons available, the Q-network has 15 possible choices. We illustrate two actions along with their resulting graph states (i.e., next states) and their corresponding Q-values as computed by the Q-network. As shown, RLGS selects the action with the highest Q-value (pointed by the black arrow) and introduces a new emitter labeled $e_1$. For all subsequent action choices, a black arrow is used to denote the option with the highest Q-value, as selected by RLGS.
Next, a receptive field $rf$ is assigned, with its size set to 8, $rf$ includes the initial emitter $e_1$ and 7 vertices that are closest to it in terms of the Manhattan distances.


In step \circledwhite{2}, a Typle-II absorption and seven Emitter Swaps are possible actions. RLGS selects the Type-II absorption as it offers the best Q-value (i.e., -10.5 as predicted by the Q-network). As a result, a photon next to the emitter $e_1$ is absorbed, and RLGS updates the graph state accordingly. Furthermore, it includes a new photon to the receptive field (notated by the purple dotted square), which is the closest non-receptive field photon to the initially assigned emitter. The remaining actions are selected in a similar manner, as illustrated by step \circledwhite{3}-\circledwhite{8} in Fig.\ref{fig:example}, with the inference process stopping once all photons and edges have been removed from the graph state.
There is an exception in step \circledwhite{5}, where an Emitter Swap is chosen by RLGS. Since the number of vertices in the receptive field remains at 8, no new photon is added.

After obtaining the action sequence, RLGS derives the \textbf{forward} generation sequence, as shown in Fig.~\ref{fig:example}(b). The gates for each action are highlighted within orange dotted squares. Using this generation sequence, we can obtain the three fidelity metrics. 
Although $e_2$ is introduced after $e_1$, their gate operations can be executed in \textbf{parallel} within the generation sequence. RLGS enables this parallelism by identifying all quantum gates that can be executed simultaneously for each emitter, thereby optimizing the generation time.

\subsection{Complexity of RLGS}
\label{sec: complexity}
We now analyze the complexity of RLGS. In the training stage, a total of $N$ episodes are executed. For a graph state with $V$ photons and $E$ edges, each photon can be swapped or absorbed by an emitter, and edges can be removed through absorption or CZ gates. Each episode consists of at most $V+E$ iterations. Thus, there are a total of $O(N\cdot (V+E))$ iterations in training. During the experience phase, an action is selected from all possible actions in an iteration, either randomly or predicted by the Q-network. 
A photon can be absorbed using any of the three absorption operations or swapped by an emitter, while an emitter can perform a CZ gate operation with other emitters. Since these actions depend on the number of photons and emitters in the graph state, the total number of possible actions is proportional to $V$. 
In the training phase, a batch of $B$ transitions is sampled from the replay buffer, and the Q-network is updated for each transition. For each transition, the target Q-network goes through $V$ possible actions in the worst case. This results in a total $O(V+B \cdot V)$ complexity for each iteration. 
Consequently, the complexity of the RLGS training is $O(N \cdot (V+E) \cdot (V+B \cdot V))$. 

Similar to the training, the inference has at most $V+E$ iterations, with each iteration offering $V$ possible actions. However, using a receptive field restricts actions to vertices within it, reducing the number of possible actions from $V$ to $W$. Thus, the overall complexity of the RLGS inference is $O((V+E) \cdot W)$. Even if the receptive field size equals the photon count $V$ of the input graph state (i.e., the maximum receptive field size), the complexity becomes $O((V+E) \cdot V)$, which is better than the complexity of the Stabilizer Solver $O(V! \cdot V^4)$~\cite{baseline0}. Although the complexity of training is higher than that of inference, it only needs to be performed once.


\section{Evaluation}
\label{sec: evaluate}
\subsection{Experiment Setup}
\label{sec:setting}
\subsubsection{Benchmark} We use six representative quantum applications to evaluate RLGS: hardware efficient ansatz ({\tt hwea})~\cite{hwea}, linear hydrogen atom chain ({\tt hc})~\cite{hc}, Quantum Fourier Transform ({\tt qft})~\cite{qft}, Bernstein–Vazirani ({\tt bv})~\cite{bv}, 3-regular Quantum Approximate Optimization Algorithm ({\tt qaoa})~\cite{qaoa}, and quantum supremacy ({\tt supre})~\cite{supre}. For each quantum application,  we evaluate different number of logical qubits (from small to large) which lead to different graph state sizes: The \textit{small} graph states contain fewer than 50 photons, the \textit{medium} graph states contain more than 100 photons, and the \textit{large} graph states contain more than 200 photons. 
The detailed information, including the logical qubit count, number of vertices ($V$), and number of edges ($E$) are shown in Table~\ref{table:benchmark}. 

\vspace{-5pt}



\begin{table}[h!]
\setlength{\tabcolsep}{2.5pt}
    \caption{Information for each benchmark.}
    \vspace{-5pt}
  \scriptsize 
  \centering
  \begin{tabular}{|c|c|c|c|c|c|}

    \hline
    \textbf{Application} & \textbf{Benchmark}& \textbf{Logical qubit count}& \textbf{\textit{V}} & \textbf{\textit{E}}& \textbf{Graph state size}\\
    \hline
     & {\tt hwea-6} & 6& 26 & 25& \textit{small}\\
    \cline{2-6}
    {\tt hwea} &{\tt hwea-30} & 30& 122 & 121& \textit{medium}\\
    \cline{2-6}
    & {\tt hwea-52} & 52& 210& 209& \textit{large}\\
    \hline
    &{\tt hc-6} & 6& 36 & 45& \textit{small}\\
    \cline{2-6}
    {\tt hc} & {\tt hc-18} & 18& 108 & 243& \textit{medium}\\
    \cline{2-6}
    & {\tt hc-38} & 38& 228 & 893 & \textit{large}\\
    \hline
    & {\tt qft-5}  & 5& 35 & 40& \textit{small}\\
    \cline{2-6}
    {\tt qft} & {\tt qft-10}  & 10& 121 & 211 & \textit{medium}\\
    \cline{2-6}
    & {\tt qft-14}  & 14& 235& 366 & \textit{large}\\
    \hline
    & {\tt bv-6}  & 6& 18 & 17& \textit{small}\\
    \cline{2-6}
    {\tt bv} & {\tt bv-34}  & 34& 102 & 101 & \textit{medium}\\
    \cline{2-6}
    & {\tt bv-68}  & 68& 204& 203& \textit{large}\\
    \hline
    &{\tt qaoa-6}  & 6& 42 & 54& \textit{small}\\
    \cline{2-6}
    {\tt qaoa} &{\tt qaoa-18}  & 18& 130 & 166& \textit{medium}\\
    \cline{2-6}
    & {\tt qaoa-30}  & 30& 216 & 276 & \textit{large}\\
    \hline
    & {\tt supre-6}  & 6& 25 & 25& \textit{small}\\
    \cline{2-6}
    {\tt supre} & {\tt supre-26}  & 26& 112 & 111 & \textit{medium}\\
    \cline{2-6}
    & {\tt supre-52} & 52& 233 & 233 & \textit{large}\\
    \hline
  \end{tabular}
  \label{table:benchmark}
\vspace{-10pt}
\end{table}

\subsubsection{Baseline} We use Stabilizer Solver as our {\bf baseline}. To avoid an exhaustive search of all possible emission orders (Section~\ref{sec: solver}), we run the Stabilizer Solver 100 times, each with a random emission order. Among the 100 runs, we identify and report the generation sequence that yields the best result (in terms of the three metrics).


\subsubsection{Metrics} We use the three fidelity metrics described in Section~\ref{sec:error}: i) generation time $T_{gen}$, ii) number of emitters $N_e$, and iii) number of CZ gates $N_{CZ}$. We report the reduction ratio of the three metrics compared to the baseline. 
We use the parameters of quantum dot emitters for evaluation as they are notable for producing high-quality photons~\cite{quantumdot, repeater}. The parameters are as follows: emission (emitter-photon CNOT gate) time is 0.1ns~\cite{russo2018photonic}, single-qubit gate time is 0.1ns~\cite{zhan2023performance}, CZ gate time is 10ns~\cite{QDreview}, emitter coherence time $T_2$ is 4.4$\mu$s~\cite{coherence}, and single CZ gate fidelity $\sigma_{CZ}$ is 99\%~\cite{russo2018photonic, QDreview}.

\subsubsection{RLGS setting} For the Q-network, we use two Graph Isomorphism Network (GIN) layers~\cite{xu2018powerful} followed by three MLP layers. Each layer, including both GIN and MLPs, has a hidden dimension of 128. This architecture efficiently captures the structural features of the graph state while maintaining a manageable network size. 
We set the initial exploration parameter $\epsilon$ as 1 and the decay rate $\delta$ to 0.99.
Additionally, we set the discount rate $\gamma$ as 0.99 to favor the long-term expected reward.
We set the receptive field size $W$ as half of the photon number $V$, achieving a balance between computational complexity and fidelity. Additionally, we set the reward parameter $\alpha$ in Equation~\ref{eq:reward} to 0.5, which balances the trade-off between minimizing generation time and the other two metrics.
A more detailed analysis of different choices for $W$ and $\alpha$ will be presented in the sensitivity study in Section~\ref{sec: sensitivity}. Additionally, we set the rest of the parameters of RLGS as follows: training episode $N=300$, replay buffer size $M=10,000$, batch size $B=256$, and target update frequency $C=500$ iterations.


\textbf{RLGS training. }We offline train RLGS using only a subset of small benchmarks. Specifically, we choose three from the six small benchmarks and use them to train RLGS. In the main results in Section~\ref{sec: main result}, we choose {\tt hwea-6}, {\tt hc-6}, and {\tt qft-5}. We also report the results using different combination of small benchmarks for training in Section~\ref{sec: cross}. Note that, using small benchmarks in training significantly reduces the overheads involved in RL and also yields good results in all sized benchmarks as we elaborate later. 

\textbf{RLGS inference. }After offline training, we evaluate RLGS on all the 18 benchmarks in Table~\ref{table:benchmark}.

\begin{figure*}
    \centering
   \includegraphics[width=1\linewidth]{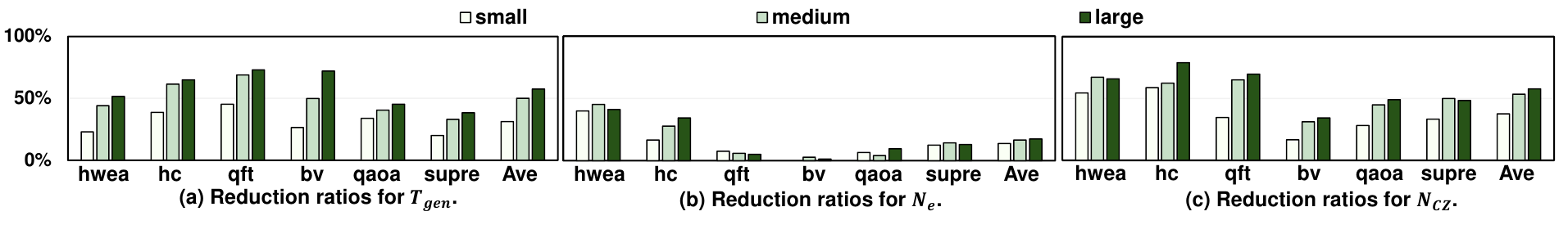}
   \vspace{-25pt}
     \caption{Reduction ratios achieved by RLGS.}
    \label{fig:main}
 \vspace{-5pt}
 \end{figure*}

\subsection{Results} 
\label{sec: main result}

As shown in Fig.~\ref{fig:main}, on average, RLGS achieves a reduction of 31.1\%, 49.6\%, and 57.5\% in $T_{gen}$ for \textit{small}, \textit{medium}, and \textit{large} benchmarks, respectively. Similarly, the reduction ratio for $N_e$ is 13.9\%, 16.7\%, and 17.5\%, while $N_{CZ}$ is 37.7\%, 53.4\%, and 57.8\%. Without specified otherwise, the results in the evaluation sections are obtained by training the RLGS using {\tt hwea-6}, {\tt hc-6}, and {\tt qft-5} and inference on all 18 benchmarks from Table~\ref{table:benchmark}. Results of training using other benchmarks are given in Section~\ref{sec: cross}
.
This result demonstrates RLGS's effectiveness in identifying graph state generation sequences that optimize the three metrics. Notably, RLGS outperforms the baseline for all metrics regardless of the benchmarks used for training or testing, showing its robustness and generalization across different application sizes and types. 

The averaged reduction ratios for all three metrics increase progressively from the \textit{small} to the \textit{large} benchmarks. This is because RLGS demonstrates robustness and consistency in handling the factorially increasing search space as graph state sizes grow. In contrast, the baseline method (i.e., the Stabilizer Solver) relies on an exhaustive search to identify the optimal generation sequence, making it less efficient with larger benchmarks.
The reductions in $T_{gen}$ and $N_{CZ}$ are more significant than the reduction in $N_{e}$. 
This is because RLGS employs a reward function that jointly optimizes all three metrics while the baseline only minimizes $N_{e}$. 
Despite this, RLGS consistently achieves fewer $N_{e}$ than the baseline for all benchmarks, demonstrating RLGS's effectiveness in optimizing any metric in a vast search space.

\begin{figure}
    \centering
   \includegraphics[width=1\linewidth]{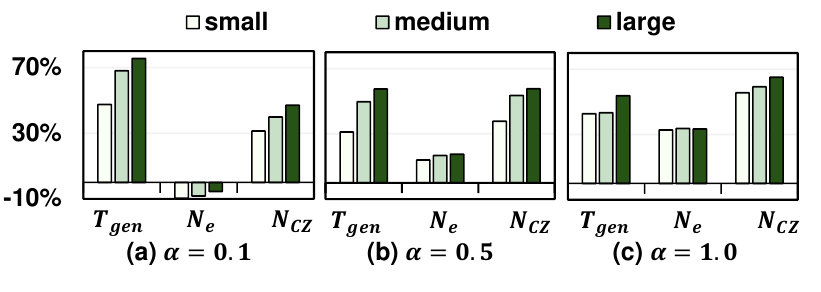}
   \vspace{-25pt}
     \caption{Results comparison with a different $\alpha$ values.}
    \label{fig:alpha}
 \vspace{-15pt}
\end{figure}

 \begin{figure}
    \centering
   \includegraphics[width=1\linewidth]{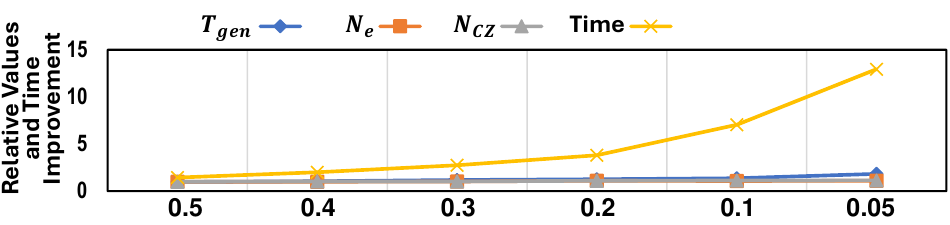}
   \vspace{-15pt}
     \caption{Relative values for different receptive field sizes.}
    \label{fig:window}
\vspace{-15pt}
 \end{figure}

  \begin{figure}
    \centering
   \includegraphics[width=1\linewidth]{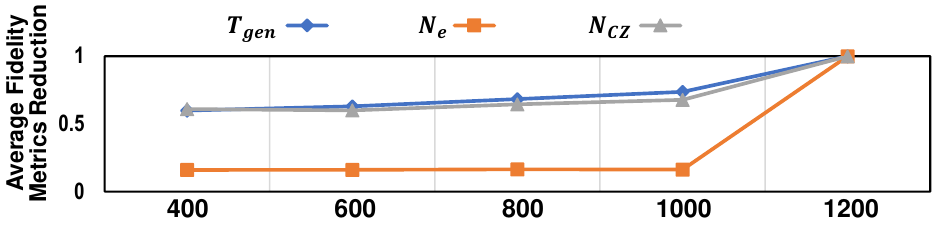}
   \vspace{-10pt}
     \caption{Metrics reduction for large graph states.}
    \label{fig:scale}
\vspace{-15pt}
 \end{figure}
 
 \begin{figure*}
    \centering
   \includegraphics[width=1\linewidth]{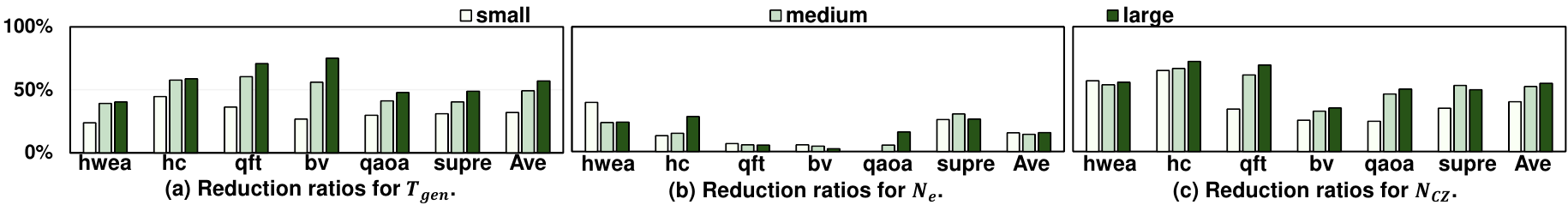}
   \vspace{-25pt}
     \caption{Averaged reduction ratios over 20 Q-networks.}
    \label{fig:c63}
 \vspace{-13pt}
 \end{figure*}

\subsection{Sensitivity Study}
\label{sec: sensitivity}
\subsubsection{Reward parameter $\alpha$}
As introduced in Section~\ref{sec:formulation}, we use a parameter $\alpha$ to adjust the penalty for the number of emitters. This allows us to balance the $T_{gen}$ with $N_{CZ}$ and $N_e$. In addition to using $\alpha=0.5$ in our main result, we also evaluate the impact of setting $\alpha$ to 0.1 and 1. We present the average reduction ratios of $T_{gen}$, $N_e$, and $N_{CZ}$ for $\alpha=0.1$, $\alpha=0.5$ and $\alpha=1$, as shown in Fig.~\ref{fig:alpha}.

When $\alpha=0.1$, the penalty for applying an Emitter Swap is reduced, leading RLGS to focus more on minimizing $T_{gen}$. As shown in Fig.~\ref{fig:alpha}(a), the reduction ratios for $T_{gen}$ improve to 47.4\%, 68.1\%, and 75.5\% compared to the baseline for the \textit{small}, \textit{medium}, and \textit{large} benchmarks, respectively. However, $N_e$ are slightly worse than the baseline, while $N_{CZ}$ reduction ratios are 31.4\%, 39.9\%, and 47.0\% for \textit{small}, \textit{medium}, and \textit{large} benchmarks, respectively. This result shows that RLGS prioritized a shorter graph state generation time at the cost of using more emitters and more CZ gates when $\alpha=0.1$.   

When $\alpha=1.0$, RLGS places a higher focus on reducing both $N_{e}$ and $N_{CZ}$. As shown in Fig.~\ref{fig:alpha}(c), the reduction ratios for $T_{gen}$ are 42.6\%, 43.6\%, and 53.8\% for \textit{small}, \textit{medium}, and \textit{large} benchmarks, respectively. 
However, the $N_e$ reduction ratios increase to 32.8\%, 33.7\%, and 33.4\%, while the reduction ratios of $N_{CZ}$ increase to 55.6\%, 59.3\%, and 65.1\% for the \textit{small}, \textit{medium}, and \textit{large} benchmarks. These results indicate that RLGS shifts its focus towards reducing $N_e$ and $N_{CZ}$ by sacrificing some optimization in $T_{gen}$. Interestingly, we observe an improvement in $T_{gen}$ reduction ratio for \textit{small} benchmarks compared to $\alpha=0.5$. This is because smaller benchmarks require fewer emitters, resulting in limited opportunities for parallel CZ gate execution. For example, when $N_e=2$ in the example of Fig.~\ref{fig:example}, no parallelism is possible for CZ gates. In \textit{small} benchmarks, reducing $N_{CZ}$ directly decreases the critical path depth, thereby reducing $T_{gen}$. However, for larger benchmarks, the higher penalty reduces the number of emitters, which in turn limits the parallelism of CZ gates and results in increased $T_{gen}$.


\subsubsection{Receptive Field Size}
\label{sec: window}
To evaluate the impact of the receptive field size $W$ on the three metrics, we conduct evaluations using six receptive field sizes: $0.5V$, $0.4V$, $0.3V$, $0.2V$, $0.1V$, and $0.05V$, where $V$ is the number of photons in the graph state. We test each receptive field size using all \textit{large} benchmarks and compare these results to $W=V$, as shown in Fig.~\ref{fig:window}. Also, we show the runtime improvement for each receptive field size to determine the trade-offs between optimization effectiveness and computational efficiency.

When the receptive field size is $0.5V$, the values of $N_e$ and $N_{CZ}$ are nearly identical to those in $W=V$. However, $T_{gen}$ increases by 4.0\%, while the runtime is $1.45\times$ faster compared to $W=V$. Although the $T_{gen}$ is slightly affected, using a receptive field size $0.5V$ improves computational efficiency. Given this balance between performance and speed, we set $W=0.5V$ as our default setting.

As the receptive field size $W$ decreases, both $N_e$ and $N_{CZ}$ gradually increase, reaching up to $1.4\times$ compared to $W=V$ when $W=0.05V$. This is because a smaller receptive field size restricts the selection of actions, which forces RLGS to perform additional Emitter Swaps, leading to more $N_e$ and $N_{CZ}$, as discussed in Section~\ref{sec:formulation}.
Moreover, as the receptive field size $W$ decreases, $T_{gen}$ increases, reaching up to $1.82\times$ compared to $W=V$ when $W=0.05V$. This occurs because a larger receptive field size allows emitters to swap with photons that are not nearby, enabling better distribution of emitters and parallel execution. In contrast, a smaller receptive field size restricts Emitter Swaps to adjacent photons, limiting parallelism and resulting in longer generation times. 
In return, reducing the receptive field size to $W=0.05V$ results in a significant runtime speedup of up to $13.0\times$ compared to using $W=V$.      

\subsection{Scalability}
To evaluate the scalability of RLGS, we compared it against the baseline using five additional graph state sizes: 400, 600, 800, 1000, and 1200 photons. 
We compare the reduction ratios across the three metrics for each size, as shown in Fig.~\ref{fig:scale}. As observed, both the Stabilizer Solver and RLGS are able to generate at least one valid generation sequence for graph states with fewer than 1200 photons. However, while the Stabilizer Solver fails to find a solution for any graph states with 1200 photons, RLGS successfully identifies a generation sequence for every graph state at this size. Therefore, we set the reduction ratios as 1 for all three metrics in 1200-photon graph states. From 400 photons to 1000 photons, we observe a consistent increase in the reduction ratio for $T_{gen}$ and $N_{CZ}$, while the reduction ratios for $N_e$ remain constant. This is because, as the graph state size increases, the Stabilizer Solver identifies fewer valid generation sequences. It struggles to find sequences with shorter $T_{gen}$ and fewer $N_{CZ}$. However, as $N_e$ is its main optimization goal, it maintains a relatively low $N_e$ across all photon counts.

\subsection{Training RLGS Using Different Benchmarks}
\label{sec: cross}
We further evaluate the robustness and generalization of RLGS by training Q-networks on all combinations of three \textit{small} benchmarks and evaluate the resulting Q-networks across all the benchmarks in Table~\ref{table:benchmark}. With six \textit{small} graph states, this results in ${6\choose 3}=20$ possible training combinations and 20 trained Q-networks. 
The results are shown in Fig.~\ref{fig:c63}. On average, RLGS achieves reductions in $T_{gen}$ of 32.1\%, 49.1\%, and 57.0\% for \textit{small}, \textit{medium}, and \textit{large} graph states, respectively. Reductions in $N_e$ are 14.9\%, 13.8\%, and 15.0\%, while reductions in $N_{CZ}$ is 39.9\%, 52.0\%, and 54.7\%. 

Compared to the results in Fig.~\ref{fig:main}, which are derived from a single trained Q-network, we observe that the three metrics are generally slightly better than that in Fig.~\ref{fig:c63} for applications where \textit{small} graph states were included in the training set. For example, in Fig.~\ref{fig:main}, the Q-network is trained using the \textit{small} {\tt hwea}, {\tt hc}, and {\tt qft}. For these applications, most of the three metrics reported in Fig.~\ref{fig:main} are equal to or slightly better than those in Fig.~\ref{fig:c63}. On the other hand, for graph states from the remaining three applications ({\tt bv}, {\tt qaoa}, and {\tt supre}), most of the three metrics in Fig.~\ref{fig:c63} are equal or slightly better than those in Fig.~\ref{fig:main}. This observation suggests that Q-networks trained on specific graph states show slight improvements when tested on related applications. Meanwhile, RLGS maintains robust performance for all unseen applications.

\section{Related Works}
There are several pioneering compilation frameworks for PQCs~\cite{gs1, gs2, li2023minimizing, li2023orchestrating, mo2024fcm, oneq, oneperc}. Among them, \cite{li2023minimizing, li2023orchestrating} focus on minimizing the depth of a photonic cluster state to improve fidelity, whereas photonic cluster state is another type of PQC that leverages a 2-D lattice entangled photon. Additionally, \cite{gs1, gs2} focus on reducing the size of a graph state for a quantum circuit, which allows us to use a smaller graph state to achieve the same quantum application. However, they do not focus on emitter-based graph state generation. Moreover, \cite{oneperc, oneq, mo2024fcm} aims to reduce the impact of fusion failures for fusion-based graph state generation. However, none of these works focus on identifying the generation sequence for the emitter-based photonic graph state generation scheme.

\section{Conclusion}
In this paper, we propose RLGS, a novel approach to identify the generation sequence of graph state with optimized three key metrics. RLGS leverages Reinforcement Learning (RL) and Graph Neural Network (GNN) to efficiently identify the optimal graph state generation sequence. 
Experimental results show that RLGS achieves an average reduction in generation time of 31.1\%, 49.6\%, and 57.5\% for small, medium, and large graph states compared to the baseline. 
Moreover, the reductions in the number of quantum emitters are 13.9\%, 16.7\%, and 17.5\%, whereas the reductions in the number of CZ gates are 37.7\%, 53.4\%, and 57.8\%, respectively.

\end{document}